\documentclass[conference]{IEEEtran}
\IEEEoverridecommandlockouts
\usepackage{cite}
\usepackage{amsmath,amssymb,amsfonts}
\usepackage{algorithmic}
\usepackage{graphicx}
\usepackage{textcomp}
\usepackage{xcolor}
\usepackage{booktabs} 
\usepackage{listings}
\usepackage{multirow}
\usepackage{soul}
\usepackage{xspace}
\usepackage{graphicx}
\usepackage{ifthen}
\usepackage[normalem]{ulem} 
\usepackage{xcolor,colortbl}
\usepackage{hyperref}

\newboolean{showedits}
\setboolean{showedits}{true} 
\ifthenelse{\boolean{showedits}}
{
	\newcommand{\del}[1]{\textcolor{red}{\sout{#1}}} 
	\newcommand{\nbe}[3]{
		{\colorbox{#3}{\bfseries\sffamily\scriptsize\textcolor{white}{#1}}}
		{\textcolor{#3}{\sf\small$\blacktriangleright$\textit{#2}$\blacktriangleleft$}}}
}{
	\newcommand{\del}[1]{} 
	
	\newcommand{\nbe}[3]{}
}

\newboolean{showcomments}
\setboolean{showcomments}{true} 
\newcommand{\id}[1]{$-$Id: scgPaper.tex 32478 2010-04-29 09:11:32Z oscar $-$}

\ifthenelse{\boolean{showcomments}}
 {
 	\newcommand{\nbc}[3]{
 		{\colorbox{#3}{\bfseries\sffamily\scriptsize\textcolor{white}{#1}}}
		{\textcolor{#3}{\sf\small$\blacktriangleright$\textit{#2}$\blacktriangleleft$}}}
	
 }{
 	\newcommand{\nbc}[3]{}
 	
 }

\usepackage[most]{tcolorbox}
\ifthenelse{\boolean{showedits}}
{
  \newtcolorbox{inserted}{%
       title=Inserted text:,
       colframe=blue,colback=blue!5!white,
       breakable,
       leftrule=0mm, 
       bottomrule=0mm,
       rightrule=0mm,
       toprule=0mm,
       arc=0mm, outer arc=0mm,
       oversize
  }
  \newtcolorbox{deleted}{%
       title=Deleted text:,
       colframe=red,colback=red!5!white,
       breakable,
       leftrule=0mm, 
       bottomrule=0mm,
       rightrule=0mm,
       toprule=0mm,
       arc=0mm, outer arc=0mm,
       oversize
  }
  \newtcolorbox{refactored}{%
       title=Rewritten text:,
       colframe=blue,colback=red!5!white,
       breakable,
       leftrule=0mm, 
       bottomrule=0mm,
       rightrule=0mm,
       toprule=0mm,
       arc=0mm, outer arc=0mm,
       oversize
  }
}{

}
\newboolean{isblinded}
\setboolean{isblinded}{true}
\ifthenelse{\boolean{isblinded}}
{\newcommand\blind[1]{BLINDED\xspace}}
{\newcommand\blind[1]{#1\xspace}}

\newcommand{\commented}[1]{}

\newcommand{\eg}{\emph{e.g.,}\xspace}
\newcommand{\ie}{\emph{i.e.,}\xspace}
\newcommand{\etal}{\emph{et al.}\xspace}

\definecolor{source}{gray}{0.95}
\usepackage[T1]{fontenc} 

\lstdefinelanguage{Java}{
  tabsize=4
}[keywords,comments,strings]

\definecolor{source}{gray}{0.95}
\definecolor{highlight}{gray}{0.9}

\lstnewenvironment{codesnippet}{%
	\lstset{%
		frame=single,
		framerule=0pt,
		mathescape=false
	}
}{}

\newcommand{\SO}{Stack Overflow\xspace}

\newcommand{\rqone}{What types of crypto challenges do developers face in cryptography?\xspace}

\begin{document}

\title{Hurdles for Developers in Cryptography}


\author{\IEEEauthorblockN{Mohammadreza Hazhirpasand}
\IEEEauthorblockN{Oscar Nierstrasz}
\IEEEauthorblockA{University of Bern\\
Bern, Switzerland}
\and
\IEEEauthorblockN{Mohammadhossein Shabani}
\IEEEauthorblockA{Azad University\\
Rasht, Iran}
\and
\IEEEauthorblockN{Mohammad Ghafari}
\IEEEauthorblockA{School of Computer Science\\University of Auckland\\
Auckland, New Zealand}
}

\maketitle

\begin{abstract}
Prior research has shown that cryptography is hard to use for developers.
We aim to understand what cryptography issues developers face in practice.
We clustered 91 954 cryptography-related questions on the Stack Overflow website, and manually analyzed a significant sample (\ie 383) of the questions 
to comprehend the crypto challenges developers commonly face in this domain. 
We found that either developers have a distinct lack of knowledge in understanding the fundamental concepts, \eg OpenSSL, public-key cryptography or password hashing,  or the usability of crypto libraries undermined developer performance to correctly realize a crypto scenario.
This is alarming and indicates the need for dedicated research to improve the design of crypto APIs.
\end{abstract}


\section{introduction}
Studies have shown that cryptography concepts are hard to understand for developers, and the complexity of crypto APIs has rendered their secure usage very difficult \cite{nadi2016jumping} \cite{hazhirpasand2020java}.
There exist static analysis tools, but developers are reluctant to employ them due to a lack of familiarity, restrictions in organizational policies, or high rates of false positives \cite{Tymchuk2018, Corrodi2018}.
Researchers have recently developed new APIs to ease the adoption of cryptography \cite{kafader2021}, yet online Q\&A forums are among the main information sources used to resolve developer issues.


Closer inspection of online forums such as Stack Overflow provides a shortcut to identifying the frequent challenges that developers face in this domain.
Therefore, 
in this study, we address the following research question: \emph{\rqone}
We extract the common problems that developers recently encounter when dealing with various areas of cryptography. 
The findings provide significant help for developers in general, and software team leaders, tutors and crypto library designers in particular, to raise their awareness of common misunderstandings, or to highlight areas with a steep learning curve.

Unlike other studies, we only focus on crypto-related challenges of developers.
To cover various types of crypto-challenges, we need to identify different groups of questions that are similar in terms of context.
Particularly, manual grouping of such a large number of questions (\ie 91 954) is a demanding task.
We therefore used the Latent Dirichlet Allocation (LDA) generative statistical model, and found three main topics in 91 954 crypto-related posts on \SO.
We then used stratified sampling to study 383 posts randomly from the three topics to identify the most common problematic issues for developers. 
The results showed that developers commonly failed to implement a cryptographic scenario due to two reasons, namely the complexity of crypto APIs, and their lack of familiarity with fundamental concepts such as digital certificates, public-key cryptography, and hashing algorithms.

Our findings show that hurdles for developers in cryptography are not yet resolved, and due to its impact on security, this domain urgently needs dedicated research effort.
We are conducting a survey with developers who actively helped the Stack Overflow community in this domain to understand potential remedies to this problem. 



\section{Related work}
\label{sec:related}
Sifat \etal investigated three online sources, \ie  Crypto Stack Exchange,  Security Stack Exchange, and Quora, to identify complications with respect to implementing security in data transmission \cite{jahan2017exploratory}.
Their findings suggest that the most discussed technique is transport layer security (TLS), and the Cross-Site Scripting (XSS) attack is the main concern of developers. 
In another study, Yang \etal conducted a large-scale analysis of security-related questions on \SO \cite{yang2016security}. 
They identified five main categories, \ie web security, mobile security, cryptography, software security, and system security but they did not look into the challenges of each topic. 
A recent study conducted by Meng \etal has recognized the challenges of writing secure Java code on \SO \cite{meng2018secure}.
Their examinations provide compelling evidence that security implications of coding options in Java, \eg CSRF tokens, are not well-perceived by a large number of developers.
Nandi \etal conducted an empirical study on the frequent crypto obstacles with which Java developers commonly face \cite{nadi2016jumping}.
They triangulated data from a survey, 100 randomly selected Java GitHub repositories, and the top 100 Java cryptography questions asked on Stack Overflow. 
Their analyses depicted nine main crypto topics, suggesting that developers face difficulties using cryptography.
This issue has adversely affected developer performance and software security \cite{fischer2019stack}.
A recent study showed that developers blindly use the provided vulnerable code snippets found on \SO \cite{fischer2017stack}.
They mentioned that 15.4\% of the 1.3 million Android applications contained security-related code snippets from \SO.
The previous studies solely focused on security or crypto implications of a particular language or in general security-related concerns.
In contrast, we specifically analyzed crypto-related questions of any kind irrespective of any programming languages or particular part of cryptography.

\section{Methodology}
\label{sec:method}
We first explain the data gathering procedure and then describe how we clean the data, and briefly introduce the LDA topic modeling.


\subsection{Data Extraction}
\label{subsec:dataext}
To collect crypto-related posts on \SO, we assumed that the attached tags to a question mainly reflect the question's topic.
We first used the ``cryptography'' tag, \ie \emph{base tag}, to fetch crypto-related posts, \ie 11\,130 posts, with the help of the Data Explorer platform (Stack Exchange).
We found 2\,184 tags (\emph{candidate tags}) that occurred in posts together with the ``cryptography'' tag.
However, not all candidate tags were crypto-related \eg C\#.

To find relevant posts with the base tag, we used two metrics to determine which of the candidate tags are exclusively related to the base tag.
We introduced the first metric as \emph{affinity} to determine the degree to which a candidate tag (T) is exclusively associated with the base tag (BT).
For each \emph{T}, we used the \emph{posts with tags} function, for brevity pwt(), to calculate the number of posts whose tags contain both \emph{T} and \emph{BT} .
We used pwt() to obtain the number of posts whose tags contain \emph{T}.
Given these two values, we compute \emph{affinity(T,BT) = $|$pwt(T,BT)$|$ / $|$pwt(T)$|$}, whose result ranges from zero to one.

The smaller the value of the first metric, the weaker the association between \emph{T} and \emph{BT}.
For example, the ``C++'' and ``encryption'' tags each appeared 639\,897 and 29\,737 times respectively in the entire \SO.
The ``C++'' tag appeared together with BT 540 times and ``encryption'' was used 3535 times with BT.
The value of affinity for the ``C++'' tag is 0.0008 and 0.1188 for the ``encryption'' tag, values which demonstrate a strong affinity for ``encryption'' and BT. 

However, higher values of affinity for some candidate tags do not necessarily indicate tags that are closely related to cryptography.
For example, the ``s60-3rd-edition'' tag appeared once with the base tag and in total 11 times in \SO.
The value of affinity for this candidate tag is 0.09, which is close to the value of the ``encryption'' tag, even though it appeared only once with the base tag.
To resolve this issue, we introduced a second metric, \emph{coverage(T,BT) = $|$pwt(T,BT)$|$ / $|$pwt(BT)$|$}.
The second metric indicates the coverage of the BT posts by T.
As an example, the value (\ie 0.00008) of coverage for the ``s60-3rd-edition'' tag proves that the candidate tag does not exclusively cover the base tag while 
 the ``C++'' tag covers 0.04 of the cryptography-related questions.
 
Two authors of this paper examined various combinations of thresholds for the two metrics, and manually reviewed the resulting tags. 
We noticed that the thresholds to collect \emph{only crypto-related tags} from the candidate tags (\ie 2 184) are the ones above the affinity: 0.025 and coverage: 0.005. 
There are 40 crypto-related tags that fall within the selected threshold domain. 
The list of crypto-related tags as well as their frequencies are available online.\footnote{\url{http://185.94.98.132/~crypto/paper\_data/tags.csv}}
Next, we again used Stack Exchange Data Explorer to extract posts containing each of the selected tags (\ie 40 tags) but not the base tag, and recorded them in CSV files, which are available online.\footnote{\url{http://185.94.98.132/~crypto/paper\_data/}}


\subsection{Data clustering via Topic Modeling}
We combined the title and body of a post in order to  create a document.
We removed duplicate post IDs in multiple CSV files, and finally obtained 91\,954 unique documents, without considering when the posts were created.
Evidently, each of the documents contained a large number of unnecessary text elements that could produce noise in the output of a topic modeling algorithm.
We preprocessed the documents in the following steps:
(1) we removed all the code blocks enclosed by the ``<code>'' tag,
(2) we removed all the HTML elements with the help of the Beautiful Soup library,\footnote{https://www.crummy.com/software/BeautifulSoup/}
(3) we removed newlines and non-alphanumeric characters,
(4) we used the NLTK package to eliminate English stop words from the documents, and finally
(5) we used the Snowball stemmer to normalize the text by transforming words into their root forms, \eg playing converts to play.
We found 269\,795 stemmed words in total.
Finally, we used the CountVectorizer class in Scikit-learn to transform the words into a vector of term/token counts to feed into a machine learning algorithm.


We used Scikit-learn,\footnote{https://scikit-learn.org/} a popular machine learning library in Python that provides a range of supervised and unsupervised learning algorithms. 
Latent Dirichlet Allocation (LDA) is an unsupervised learning algorithm based on a generative probabilistic model that considers each topic as a set of words and each document as a set of topic probabilities \cite{blei2003latent}.
LDA has been used to discover latent topics in documents in a large number of prior studies \cite{bangash2019developers} \cite{yang2016security} \cite{rosen2016mobile}.

Before training a model, LDA requires a number of important parameters to be specified.
LDA asks for a fixed \textit{number of topics} and then maps all the documents to the topics. 
The \textit{Alpha} parameter describes document-topic density, \ie higher alpha means documents consist of more topics, and generates a more precise topic distribution per document.
The \textit{Beta} parameter describes topic-word density, \ie higher beta means topics entail most of the words, and generates a more specific word distribution per topic.

The optimal values of hyperparameters cannot be directly estimated from the data, and, more importantly, the right choice of parameters considerably improves the performance of a machine learning model \cite{osman2017hyperparameter}.
We therefore used the GridSearchCV function in Scikit-learn to perform hyperparameter tuning to generate candidates from an array of values for the three aforementioned parameters, \ie \textit{Alpha}, \textit{Beta}, and the \textit{number of topics}.
As research has shown that choosing the proper number of topics is not simple in a model, an iterative approach can be employed \cite{zhao2015heuristic} to render various models with different numbers of topics, and choose the number of topics for which the model has the least perplexity. 
Perplexity is a measure used to specify the statistical goodness of fit of a topic model~\cite{blei2003latent}.
We therefore specified the number of topics from 1 to 25. 
We also used the conditional hyperparameter tuning for Alpha, which means a hyperparameter may need to be tuned depending on the value of another hyperparameter \cite{luo2016review}.  
We set \textit{alpha} = 50 / \textit{number of topics} and \textit{beta} = 0.01, following guidelines of previous research \cite{griffiths2004finding}.

Optimizing for perplexity, however, may not always result in humanly interpretable topics \cite{chang2009reading}.
To facilitate the manual interpretation of the topics, we used a popular visualization package, named pyLDAvis\footnote{https://github.com/bmabey/pyLDAvis}, in Python.
The two authors of this paper separately checked the resulting top keywords of the topics, \ie from 1 to 25, and the associated pyLDAvis visualizations to ensure that the given number of topics is semantically aligned with human judgment.

\subsection{Data analysis}
\label{sec:dataan}
We computed the required sample size for 91\,954 documents with a confidence level of 95\% and a margin of error of 5\%, which is 383 documents.
We then used stratified sampling to divide the whole population into smaller groups, called strata.
In this step, we considered each topic as one stratum, and randomly selected the documents proportionally from the different strata.
We then used thematic analysis, a qualitative research method for finding topics in text \cite{braun2006using}, to extract the frequent topics from the documents.
Two authors of the paper carefully reviewed the title, question body, and answer body of each document. 
Each author then improved the extracted topics by labeling the posts iteratively.
We then calculated Cohen’s kappa, a commonly used measure of inter-rater agreement \cite{cohen1960coefficient}, between the two reviewers.
Finally, the two reviewers compared their final labelling results, and re-analyzed the particular posts in a session where they disagreed in order to discuss and arrive at a consensus.
\section{Results and Discussion}
\label{sec:rdiscussion}

\begin{figure*}[ht]
\centering
\centerline{\includegraphics[width=0.90\linewidth,trim=4 4 4 4,clip]{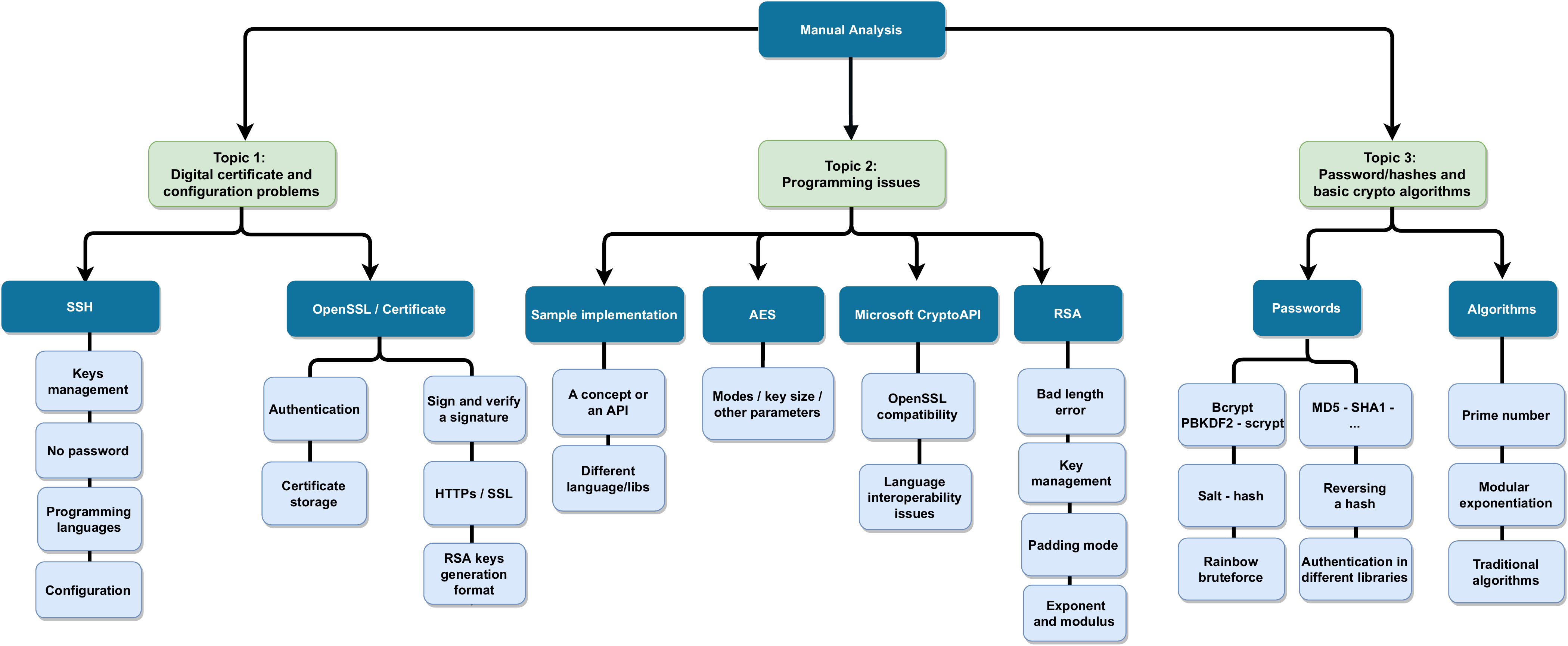}}
\caption{The results of manual analysis for the three topics}
\label{fig:tags2}
\end{figure*} 

Our hyperparameter tuning demonstrated that the best number of topics is three.
Similarly, after analyzing pyLDAvis's visualizations and top keywords for 1 to 25 topics, the two reviewers also achieved a consensus on three as the number of topics.
The pyLDAvis interactive visualization for the three topics is available online.\footnote{\url{http://185.94.98.132/~crypto/paper_data/lda.html}}
The reviewers named the topics by considering the general themes of top keywords returned by LDA (See \autoref{tab:topics}).
We determined that the first topic is about digital certificates and configuration issues, the second one is about programming issues concerning encryption and decryption, and the third concerns passwords/hashes and basic crypto-related algorithms.
As an influential indicator of topic relevancy, we realized that the frequencies of the candidate tags used in the three topics are aligned with the general themes of the topics.\footnote{\url{http://185.94.98.132/~crypto/paper_data/tags-topics.csv}}
For instance, we observed that the AES, DES, Encryption, and RSA tags are mostly used in programming issues,  the Hash, SHA, SHA256, MD5, XOR, and Salt tags are more frequent in the password/hash topic, and finally, the Digital-signature, Keystore, OpenSSL, Private-key, Public-key, Smartcard, and X509certificate tags are more common in the digital certificate topic.

With respect to stratified sampling, we considered the number of documents in each stratum (\ie each topic) as 139, 124, and 119 documents from the first topic to the third one respectively.
The selected documents were created in the last 5 years on \SO.
Extracting the themes, the reviewers achieved 79\% Kappa score, which demonstrates a substantial agreement between the two reviewers.


\begin{table}[]
\scriptsize
\caption {The three topics and their top keywords} \label{tab:topics} 
\begin{tabular}{p{2.5cm}lll} \hline
\textbf{Topic}  & \textbf{Top keywords}                                                                                                                                                                                                                                                  \\ \hline
\begin{tabular}[c]{@{}l@{}}Digital certificate and\\ configuration problems\end{tabular}    & \begin{tabular}[c]{@{}l@{}}use, certif, file, server, key, openssl, client,\\  work, tri, need, sign, user, error,applic, creat,\\ code,  secur, app,  encrypt, ssl, store, instal,\\ like, connect, problem, want, way, run, request\end{tabular}              \\    \hline
\begin{tabular}[c]{@{}l@{}}Programming issues\end{tabular}                & \begin{tabular}[c]{@{}l@{}}key, encrypt, use, decrypt, code, data, file,\\ string,  tri, public, work, ae, im, byte, need,\\ java,  generat, messag, encod, privat, rsa,\\ cipher, algorithm, block, like, implement, \\  error, problem, function, text\end{tabular} \\    \hline
\begin{tabular}[c]{@{}l@{}}Password/hashes and\\ basic crypto algorithms\end{tabular}              & \begin{tabular}[c]{@{}l@{}}hash, use, valu, password, function,like,\\ array, string,  need, code, number, key, want,\\ way, store,data, tabl, im, salt,  tri, differ,\\ time,  algorithm, work, md5, user, make, \\ generat, object, implement
\end{tabular}   \\ \hline
\end{tabular} 
\end{table}

\subsubsection{Topic One}
\textbf{Digital certificate and configuration problems.} The manual analysis for the first topic depicts that developers discussed two main areas, namely certificate/OpenSSL (63\%) and SSH (37\%).
For instance, the discussions were related to OpenSSL configuration, signing and verifying a signature, and generating PEM files using OpenSSL.
There were also questions concerning how to generate self-signed certificates, access a certificate store, create a Certificate Signing Request (CSR), establish https and secure connections, and configure certificate-based authentication in ASP.NET.
In the SSH-related questions, the majority of the users had difficulty setting SSH with no password, checking the right permission for SSH keys, using SSH programmatically, and connecting to SSH servers of other platforms (\eg Amazon).

\subsubsection{Topic Two} \textbf{Programming issues.} As for the programming issues topic, we observed that the three most frequently discussed programming languages were Java (\ie 44), C/C++ (\ie 31), and C\# (\ie 19).
In 31\% of the posts developers discussed issues related to the AES algorithm such as different encryption modes (\eg CBC and ECB) and key sizes (\eg 128, 192, and 256-bit).
In addition to symmetric encryption, 47\% of the posts were related to working with asymmetric encryption (\ie RSA).
The challenges were mostly concerned with different padding modes (\eg OAEP),  how to calculate or understand the raw modulus and exponent numbers, and how to generate and work with different key file encodings in RSA (\eg DER-encoded format, PEM, or XML).
Moreover, another evident problem was dealing with different RSA key formats, \ie Public Key Cryptography Standards (PKCS). 
The users commonly asked how to convert PKCS\#8 to PKCS\#1 or other standards, and how to programmatically generate or use different key standards in various crypto libraries (\eg Bouncy Castle). 
There were users who had problems with illegal block size errors, often misunderstanding the suitable usage of RSA, \eg encrypting a long text.
Nevertheless, the discussions were resolved by proper responses that suggested incorporating AES and RSA into the encryption/decryption scenario.
Another type of question was about the issues in Microsoft CryptoAPI (12\%). Developers reported issues on working with OpenSSL or using RSA keys from other sources, \eg  importing keys from OpenSSL into Crypto API, converting RSA keys to be used by Bouncy Castle, verifying an OpenSSL DSA signature using CryptoAPI, having extra fields in generated keys by PHP OpenSSL, and signing a message with pyOpenSSL in Python and verifying it with CryptoAPI.
Moreover, there were questions (10\%) associated with how to either implement a scenario, \eg encryption of a string with RSA public key with Swift on iOS, or deal with problems while working with more than one crypto library or programming language, \eg encryption of a string with RSA in JavaScript and decryption in Java, or decryption of a string in Java which is already encrypted using AES-256 in iOS.

\subsubsection{Topic Three} \textbf{Password/hashes and basic crypto algorithms.}  Our findings for the password/hash topic suggest that users primarily discussed problems associated with either passwords (86\%) or basic crypto algorithms (14\%).
Different facets of producing secured passwords were the topic of most discussions.
First and foremost, users were uncertain which hashing algorithms (\eg MD5, SHA-1) can provide a higher level of reliability and how password length contributes to the strength of the resulting hash.
Users lacked the required knowledge as to what salt is and how salt can maximize the security of a hash.
In addition to pointing out the pros and cons of static salt vs random salt, respondents encouraged users to use salted passwords in order to render the brute-force or the rainbow table attack prohibitively expensive. 
Developers were doubtful about which crypto functions, \ie bcrypt(), PBKDF2(), or Scrypt(), are more secure and faster, and what key differences distinguish the three functions from other hashing algorithms, \eg MD5, SHA-256.
As regards the basic crypto algorithms, users contributed to responses concerning how to produce or find prime numbers, how to use the BigInteger class for RSA modular exponentiation, how to produce unique URL safe hash or IDs, and how to solve a Caesar Cipher or substitution ciphers.
Lastly, a few users discussed how to program an authentication module in web programming frameworks such as Laravel, or CakePHP.

\subsubsection{Topic difficulty and popularity} 
We checked the popularity and difficulty level of each topic so as to determine which questions attracted more attention or received acceptable answers with a longer time span, which the same approach was used in the previous study \cite{yang2016security}. 
We used four factors to measure the popularity of a topic, namely 
the average number of views of documents, the average number of comments, the average number of favorites, and the average score of documents.
The four factors can be found in the CSV files,\footnote{\url{http://185.94.98.132/~crypto/paper_data/}} namely CommentCount, FavouriteCount, Score, and ViewCount.
We considered the average number of ViewCount as the foremost factor to judge the popularity of a topic, the question's score and the number of FavouriteCount as the second most important factors, and the average number of comments as the last factor.
To find the most difficult topic, we used two factors, namely 
the average time it takes for a document to obtain an accepted answer, and the ratio of the average number of answers in documents to the average number of the views.
We avoided recently posted questions from affecting the analysis by only including those that are older than six months. 

We infer that questions related to the usage of digital certificates, and configuration problems are the most popular (highest average ViewCount and FavouriteCount), and questions related to hashing and passwords are also viewed as popular based on the other two factors (\ie average CommentCount and Score).
From the difficulty standpoint, we notice that the programming issues topic is the most difficult topic as it had a greater average response time, and its proportion of average answers to average views is the lowest.

\subsubsection{Summary} The challenges in each theme were studied in detail to demonstrate how developers struggle to use or comprehend various areas of cryptography.
According to our findings, we believe that there are two foremost reasons with which developers mainly encounter problems in cryptography.
The first leading cause is a distinct lack of knowledge to discern \emph{why} or \emph{what} they need to use to accomplish a crypto task.
We observed ample evidence where developers lacked the confidence to choose the best algorithm or parameter, for instance, the right and safest padding option in AES.
Consequently, developers may use boilerplate code snippets from the provided answers, in spite of the answers' reliability and security.
In the second factor, although the fundamental concepts are the same, the implementation approach of a crypto concept in various crypto libraries is influential to developer performance.
Compelling evidence in findings urges that working with more than a crypto library due to using various architectures or platforms in a project creates confusion for developers regarding \emph{how} a particular problem can be resolved. 
They commonly have trouble in creating keys with one library and import them into another library or verifying a signature in a different crypto library.
Furthermore, adequate explanations and the existence of useful examples in documentations can alleviate the difficulty of using cryptography.



\section{Threats to validity}
\label{sec:threat}
In this study, we concentrate on one major platform where developers discuss crypto topics. 
This may not be sufficient as there are many other platforms, such as crypto Stack Exchange, which can provide more data to analyze.
We measured topic difficulty and popularity based on metrics used in the previous study. 
Nevertheless, these observations may not be sufficient to determine what type of crypto questions are more challenging than others.
Users may not always feel responsible for selecting a reasonable answer as an acceptable answer. 
Therefore, not having an accepted answer does not necessarily determine if the question is challenging for others.
\section{conclusion}
\label{sec:conclusion}
We conducted a large-scale study on crypto issues discussed on Stack Overflow to find out what crypto challenges users commonly face in various areas of cryptography.
Findings suggest that developers still have a distinct lack of knowledge of fundamental concepts, such as OpenSSL, asymmetric and password hashing, and the complexity of crypto libraries weakened developer performance to correctly realize a crypto scenario. We call for dedicated studies to investigate the usability of crypto APIs.
We are conducting a survey with users who actively helped the Stack Overflow community in this domain to understand the potential remedies. 

\subsection*{Acknowledgments}
We gratefully acknowledge the financial support of the Swiss National Science Foundation for the project
``Agile Software Assistance'' (SNSF project No. 200020-181973, Feb. 1, 2019 - April 30, 2022).


\bibliographystyle{IEEEtran}
\bibliography{ref}

\end{document}